# Chiral microstructures (spirals) fabrication by holographic lithography

**Yee Kwong Pang, Jeffrey Chi Wai Lee, Hung Fai Lee, Wing Yim Tam\*, C. T. Chan, and Ping Sheng**

*Department of Physics and Institute of Nano Science and Technology*
*Hong Kong University of Science and Technology*
*Clear Water Bay, Kowloon, Hong Kong, China*
*phtam@ust.hk*

**Abstract:** We present an optical interference model to create chiral microstructures (spirals) and its realization in photoresist using holographic lithography. The model is based on the interference of six equally-spaced circumpolar linear polarized side beams and a circular polarized central beam. The pitch and separation of the spirals can be varied by changing the angle between the side beams and the central beam. The realization of the model is carried out using the 325 *nm* line of a He-Cd laser and spirals of sub-micron size are fabricated in photoresist.

**OCIS codes:** (999.9999) photonic bandgap materials; (220.4000) Microstructure fabrication.

---

## 1. Introduction

Microstructures that exhibit photonic bandgap in which light with frequency inside the bandgap cannot propagate have posted many challenges to experimentalists, especially in the visible range [1, 2]. Despite progresses made in various techniques, like the self-assembly of micro-spheres [3], layer by layer micro-fabrication [4], and photo-lithography [5], it is still not easy to achieve large 3D complete bandgap. Recently, it was pointed out by Chutinan and Noda that spiral, which breaks the chiral symmetry and resembles the classical diamond structure for photonic crystals (Lin et al. in Ref. [4]), can exhibit large 3D complete bandgaps [6]. Toader and John later also took this approach by introducing a new square spiral photonic crystal [7]. Interestingly, Pendry showed recently that chiral structures can also lead to negative refraction [8]. With all of the above findings, it is highly desirable to fabricate chiral microstructures to confirm their predictions.

Two methods, the glancing angle deposition (GLAD) [9] and the multi-photon direct laser writing (DLW) [10], are well suited for fabricating chiral structures. In the GLAD method columnar structures are obtained by physical vapor deposition, in which vapor flux arrives at an oblique angle with respect to the normal of a 2D patterned substrate. The shadowing effects of the columns ensure the growth at chosen sites and by rotating the substrate during deposition spirals can be fabricated [11]. The DLW method utilizes the multi-photon absorption and pin-pointed focusing of laser light using high magnification objective to write directly into photoresist. The frequency of the laser light is chosen below the single-photon polymerization threshold of the photoresist such that no polymerization can occur at off-focus regions. However at the sharp focal point the intensity of the laser light may exceed the threshold for multi-photon absorption, leading to local polymerization of size as small as 120nm. Thus by manipulating the substrate with respect to the laser light with precision movements controlled a computer, 3D pre-designed patterns can be written directly into the photoresist [10]. Micro-sized square spirals have been fabricated recently using the DLW method [12]. Moreover, spiral structures containing metallic nano-Ag particles can also be made by the two-photon DLW method [13]. Despite the fact that large 3D bandgaps are obtained for spirals fabricated using the above two methods, there are limitations [11, 12]. In the GLAD method high quality and sub-micron size are still not easy to achieve, in addition to its low throughput and problems in large scale production. Similarly, the DLW method is limited to small sample size, ~tens of micrometers, and is rather time consuming. Furthermore, the bandgaps are in the IR range only.

It is known that holographic lithography (HL), a method combining holography and photo-induced polymerization techniques, can produce uniform periodic as well as quasi-periodic large 3D structures in photoresist in the optical range [5,14]. The HL method is very flexible and is good for fabricating various microstructures. We demonstrate in this paper the fabrication of chiral microstructures by exploiting the HL method to include circular polarized light beam as compared to linear polarized beams in previous HL studies [5]. Submicron spiral structures are fabricated in photoresist using six circumpolar linear polarized side beams and one circular polarized central beam, 6+1 configuration, from a UV light source. This method offers new opportunities for fabricating novel photon bandgap materials.

## 2. Model

Figure 1 shows the 6+1 beam configuration setup, six circumpolar side beams + one central beam, for the fabrication of chiral microstructures. The six side beams are represented by wave vectors

$$\vec{k}_n = k(\cos\frac{2(n-1)\pi}{6}\sin\varphi,\ \sin\frac{2(n-1)\pi}{6}\sin\varphi,\ \cos\varphi), \quad (1)$$

for $n = 1 - 6$. In Eq. 1 $\varphi$ is the incident angle between the side beams with the vertical (z) axis. It determines the aspect ratio $\rho = l/a$ ( $l = \frac{\lambda}{1-\cos\varphi}$ : pitch and $a = \frac{\lambda}{\sin 60^o \sin\varphi}$ : spiral separation) of the spirals. The central beam is along the z-axis given by

$$\vec{k}_0 = k(0,\ 0,\ 1). \quad (2)$$

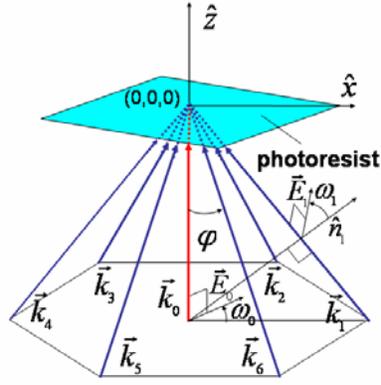

Fig. 1. The 6+1 beam configuration setup.

In Eqs. (1) and (2) $k = 2\pi/\lambda$ is the magnitude of the wave vector ($\lambda$ is the wavelength of the laser light inside the photoresist). The interference pattern, origin defined as shown in Fig. 1, of the 7 beams in Eqs. (1) and (2) is given by the following intensity profile

$$I(\vec{r}) = \sum_{n,m} \vec{E}_n e^{-i\vec{k}_n\cdot\vec{r} - i\delta_n} \cdot \vec{E}_m^* e^{i\vec{k}_m\cdot\vec{r} + i\delta_m}, \quad (3)$$

for $n, m = 0 - 6$. $\vec{E}_n$ and $\delta_n$ are the electric field and the phase of each beam, respectively. Furthermore, the electric fields of the six side beams are linear polarized with polarizations defined as the angle $\omega_n$ between the electric field $\vec{E}_n$ and the plane of incident, denoted by unit vector $\hat{n}_n$, for each beam as illustrated only for $\vec{E}_1$ in Fig. 1. The chiral property comes from the central beam which has electric field given by

$$\vec{E}_0 = \frac{E_0}{\sqrt{2}}(1,\ \pm i,\ 0). \quad (4)$$

In Eq. 4, (1, +i, 0) corresponds to right-handed chirality and (1, -i, 0) for left-handed.

In this model, φ, $\vec{E}_n$, $\omega_n$, and $\delta_n$ are the control parameters after choosing the light source. For a fixed incident angle φ and equal beam intensity, $\omega_n$ and $\delta_n$ are the only free parameters. While $\omega_n$ is easily controlled by polarizer, $\delta_n$ is more involved. It turns out that the phase of the central beam only shifts the pattern along z-axis thus does not need any adjustment. For the six phases of the side beams, four of them can be arbitrarily set to zero because it is the differences of the phases that determine the final pattern, leaving only two phases as free.

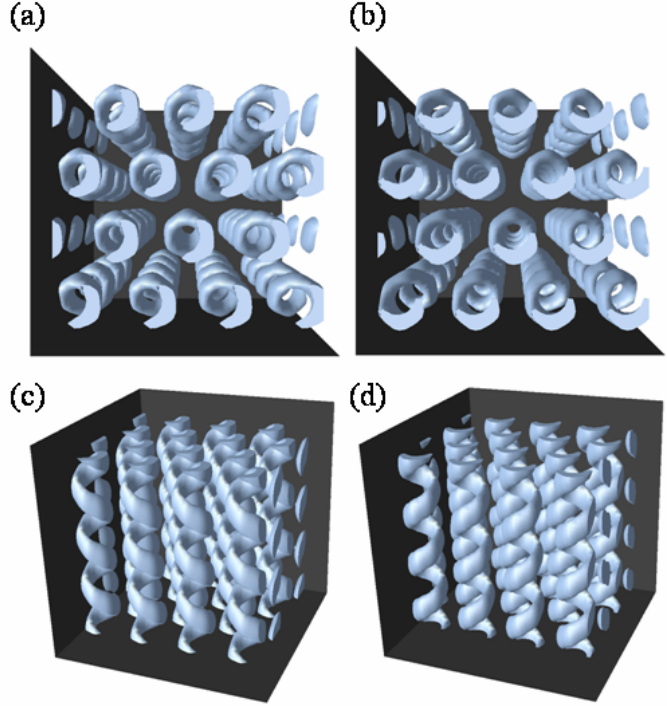

Fig. 2. Intensity contour surfaces of (a) and (c) left handed and (b) and (d) right handed spirals using the 6+1 beams interference of Eq. 3.

Figures 2(a) and (b) show the perspective top views of 3D intensity contours given by Eq. 3 using φ = 82.3°, $E_n$ = 1 and $\delta_n$ = 0 for n = 0 - 6, $\omega_0$ = 45°, and $\omega_n$ = 90° for n = 1 – 6 of well separated (a) left handed and (b) right handed spirals with aspect ratio ρ = 1.0 (l = a = 0.22 μm). Figures 2(c) and (d) show the corresponding side views of Figs. 2(a) and (b). The chiralities of the spirals are clearly shown. The spirals will touch each other with the center hole also closed up for a low enough intensity cutoff. Note that all the beams for Fig. 2 are in phase. It turns out that the quality of the spirals will be compromised when one of the side beams is out of phase with the other beams. Fortunately, good spirals can still be obtained for phase difference as large as 60° indicating that the chance of getting spiral structures is not too small. This is confirmed by 200 realizations of random-phase simulations within which 32% still shows good spirals. This ensures reasonable possibility of getting spirals in an experiment even without controlling the phases of the beams. Note that triangular "spirals" with no central hole, in contrast to the 6+1 system, can also be obtained using a 3+1 configuration [15].

### 3. Experiment

We used the 325 *nm* line of a He-Cd laser as our light source. The seven beams, 1.2 *mm* diameter, shown in Fig. 1 were obtained by splitting the laser light using a grating. The six side beams, placed symmetrically at 60° from neighboring beams around and making an angle 37.7° (corresponding to $\varphi = 21.2°$ in the photoresist) with the central axis, were linearly polarized with directions $\omega_n = 90°$ for $n = 1 – 6$ obtained by adjusting half-wave plates placed in the optical paths. The central beam, placed along the central axis as shown in Fig. 1, was converted to circular polarized using a ¼ wave plate. The seven beams were collimated to interfere at the photoresist as shown in Fig. 1. All seven beams were arranged to have roughly the same optical path without optical delays for simplicity. In principle, the phases of the beams can be adjusted by adding delay optics in the paths of the beams. Neutral density filters were used to adjust the beam power to about 80 $\mu W$ each.

We used photoresist SU8, a high contrast negative resist commonly used for near UV radiation, as the raw polymer resin. SU8 resin was dissolved in γ-butyrolactone (1: 0.7 wt.) with 2% wt. of photoinitiator Irgacure 261 (from Ciba. Co.) to form the photoresist solution. The solution was spin coated on glass substrate to form ~15$\mu m$ thick photoresist sample, after heated to ~90°C for about an hour to remove any solvent left. For optimal results, the photoresist was exposed to the interference pattern for 3$s$. A prism was placed on the top of the resin with matching fluid to lead away the interfering beams, preventing multi-reflections inside the resin during exposure. This arrangement was found crucial for fabricating good quality samples.

After the exposure, a post thermal treatment at ~90°C for 40 minutes was needed to complete the polymerization at regions where the dosage from the interference exceeded a critical value. The under-exposed un-polymerized regions were washed away by bathing the sample with propyleneglycolmethylether acetate (>2 hours) and then with acetone (~5 minutes), leaving behind a copy of the interference pattern.

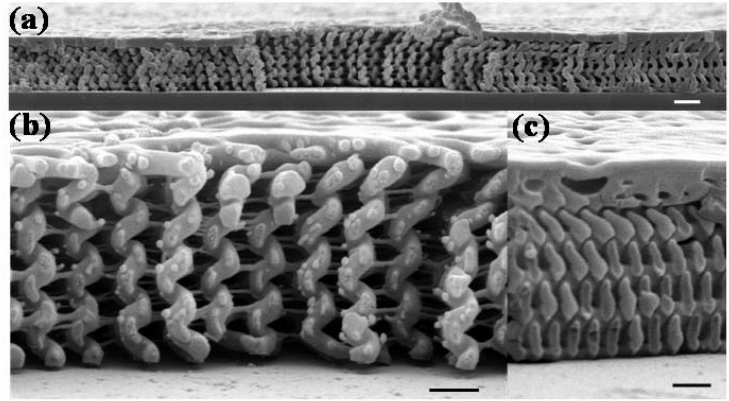

Fig.3. SEM images of spirals: (a) overall (b) close-up views. (c) Structure with out off phase interference. Scale bar is: (a) 2 $\mu m$, (b) 1 $\mu m$, and (c) 1 $\mu m$..

Figure 3 shows SEM images of columnar spirals fabricated in photoresist. It was found that the photoresist shrank substantially along the vertical direction, ~70%, at regions where spirals were observed while normal shrinkage was observed at non-spiral regions. This may be due to the stresses created in the spirals during the washing processes, in addition to the known shrinkage of the photoresist. The aspect ratio of the spirals $\rho$ is about 1.1 ($l \sim 0.9$ μm and $a \sim 0.8$ μm), much less than $\rho = 4.7$ ($l = 2.9$ μm and $a = 0.62$ μm) as given by the model. Because of the non-uniformity of the laser beams, there are regions where the phases do not favor the formation of spirals as seen at the right hand side of Fig. 3(a) and also in Fig. 3(c) from another sample. The spirals resemble the model well, shown in Fig. 1, if we allow for a shrinking of the vertical scale of the model by 70%. However, the spiral regions are unfortunately too small, ~ 20μmX20μm, and the quality of the spirals is not good enough to warrant any optical measurement. While this work is a "proof-of-principle" experiment to demonstrate the creation of spiral structures using one-phonon interference, it is conceivable that better beam quality and control of beam phases with real time monitoring of the interference pattern can give larger high quality samples to test the recently predictions of polarization gap in spiral structures [6-8, 16]. We note that the refractive index of the SU8 photoresist, ~1.62, is small and it would be highly desirable to increase the contrast so as to realize interesting effects due to the chirality. One possible way to increase the contrast is to employ the present samples as templates, e.g. in chemical vapor deposition, to fabricate higher contrast chiral microstructures."

## 4. Conclusion

We have fabricated spiral microstructures by holographic lithography using a six linearly polarized side beams and one circular polarized central beam setup. Model simulations of both right and left handed spiral microstructures can be obtained for a wide range of parameters. Realizations of the spiral microstructures are obtained in photoresist using a UV holographic setup. However, the samples are not large enough for optical measurements. Better beam profile and phase control are needed to fabricate good quality spirals in a large area for testing the theoretical predictions.

**Acknowledgments**

Support from Hong Kong RGC grants CA02/03.SC01 and HKUST603303 is gratefully acknowledged. We thank H. W. Tsang for technical help.